\documentclass{article}

\usepackage{PRIMEarxiv}

\usepackage[utf8]{inputenc} 
\usepackage[T1]{fontenc}    
\usepackage{hyperref}       
\usepackage{url}            
\usepackage{booktabs}       
\usepackage{amsfonts}       
\usepackage{nicefrac}       
\usepackage{microtype}      
\usepackage{lipsum}
\usepackage{fancyhdr}       
\usepackage{graphicx}       
\graphicspath{{media/}}     

\title{The Role of High-Performance GPU Resources in Large Language Model Based Radiology Imaging Diagnosis

}

\author{
  Jyun-Ping Kao \\
  Graduate Institute of Biomedical Electronics and Bioinformatics \\
  National Taiwan University \\
  Taiwan\\
  \texttt{jjpkao@gmail.com} \\
}

\begin{document}
\maketitle

\begin{abstract}
Large-language models (LLMs) are rapidly being applied to radiology, enabling automated image interpretation and report generation tasks. Their deployment in clinical practice requires both high diagnostic accuracy and low inference latency, which in turn demands powerful hardware. High-performance graphical processing units (GPUs) provide the necessary compute and memory throughput to run large LLMs on imaging data. We review modern GPU architectures (e.g. NVIDIA A100/H100, AMD Instinct MI250X/MI300) and key performance metrics of floating-point throughput, memory bandwidth, VRAM capacity. We show how these hardware capabilities affect radiology tasks: for example, generating reports or detecting findings on CheXpert and MIMIC-CXR images is computationally intensive and benefits from GPU parallelism and tensor-core acceleration. Empirical studies indicate that using appropriate GPU resources can reduce inference time and improve throughput. We discuss practical challenges including privacy, deployment, cost, power and optimization strategies: mixed-precision, quantization, compression, and multi-GPU scaling. Finally, we anticipate that next-generation features (8-bit tensor cores, enhanced interconnect) will further enable on-premise and federated radiology AI. Advancing GPU infrastructure is essential for safe, efficient LLM-based radiology diagnostics.
\end{abstract}


\section{Introduction}
Transformer-based Large-language models (LLMs) have shown near-human performance on complex language tasks and are now extending to medical domains. In radiology, LLMs can process both text and images, effectively simulating a radiologist's diagnostic workflow \cite{1,2, kao2025large}. For example, multimodal LLMs can generate draft radiology reports from chest X-rays by aligning visual features with the language model's embeddings \cite{kao2025large, 3}. Automation of report writing is needed: manual report generation is time-consuming, expert-intensive, and prone to variability and error. The potential of LLMs to improve productivity and reduce errors has therefore garnered strong interest \cite{1,2}.

However, translating LLM capabilities into clinical practice poses stringent requirements. Radiology Artificial Intelligence (AI) must maintain extremely high accuracy to avoid misdiagnosis and operate with low latency that the results are available during patient care \cite{4}. Emergency settings and high-volume centers may generate hundreds of images per hour, so inference delays of even seconds per case can accumulate. For instance, Salam et al. \cite{4} observed that open-source LLMs required ~6$\pm$2 seconds per report, whereas cloud-based GPT-4 took ~13$\pm$4 seconds. These timings highlight that inference speed matters. Achieving both high accuracy and low latency in LLMs requires large model capacity and fast computation. This naturally drives demand for high-performance Graphical Processing Units (GPU) hardware \cite{5,6}. 
Modern GPUs excel at parallel linear algebra, making them ideal for the large matrix multiplies in LLMs. Indeed, progress in radiology AI has been tightly coupled with GPU advances. A recent review noted that large medical imaging datasets and improved hardware (especially GPUs and distributed systems) have accelerated AI deployment in radiology \cite{5}. In short, LLM-based radiology applications are emerging due to advances in neural models, but their clinical utility hinges on fast, reliable inference. Thus, understanding GPU architectures and performance metrics is key to realizing LLM diagnostics in practice.

\section{GPU Architectures and Performance Metrics}
\label{sec:headings}

GPUs are specialized processors for massive parallel computation. Modern data-center GPUs, such as NVIDIA's Ampere (A100) and Hopper (H100) families, or AMD's Instinct series (e.g. MI250X, MI300), combine thousands of parallel cores with high-bandwidth memory. We briefly describe their relevant characteristics and the metrics that govern LLM performance:
\subsection{Compute Throughput (FLOPS)} 
Measured in floating-point operations per second (FLOPS), this indicates how many arithmetic ops a GPU can perform. Key precisions include FP32 (single precision), FP16/BF16 (16-bit half-precision or bfloat16), and lower (e.g. FP8, INT8). For example, the NVIDIA A100 GPU delivers ~19.5 teraflops (TFLOPS) at FP32 and up to 312 TFLOPS using FP16 with sparsity \cite{6}. Its successor, the H100, roughly triples the per-core throughput across data types; it supports a new FP8 precision and custom ``Transformer Engine'' that can accelerate AI workloads up to 9$\times$ faster in training and ~30$\times$ faster in inference (with large LLMs) compared to A100. AMD's Instinct MI250X (CDNA2) provides ~0.38 PFLOPS FP16 (128 GB HBM2e) and with structured sparsity can reach >1.3 PFLOPS \cite{7}. The next-gen MI300 (CDNA3) adds matrix cores supporting mixed BF16/FP8 formats, further increasing throughput.
\subsection{Memory Bandwidth and Capacity}
LLMs have massive parameter counts and activation tensors, so memory is a bottleneck. The throughput at which data can be read/written (memory bandwidth) often limits performance. A100's 40--80 GB of HBM2 memory offers ~1.6 TB/s bandwidth \cite{6}. H100's 80 GB of HBM3 nearly doubles this to over 3 TB. High bandwidth allows larger batches and faster weight fetch, improving throughput. Video RAM (VRAM) capacity also sets limits: a Transformer with millions of parameters (e.g. 175B) may need hundreds of GB unless quantized or sharded. Thus, GPUs with large VRAM (A100: 80 GB; MI300: up to 192 GB) enable bigger models or batches. Insufficient VRAM forces smaller batch size or partitioning, increasing per-sample latency.

\subsection{Tensor Cores and Mixed Precision}
Modern GPUs include specialized ``tensor cores'' designed for deep learning. These units perform matrix-multiply-accumulate extremely fast in lower precision. For instance, NVIDIA's tensor cores accelerate FP16/BF16 ops and now FP8. The H100's FP8 tensor cores ``halve data storage and double throughput'' compared to FP16/BF16 \cite{6}. This enables 2--4$\times$ speedups for LLM inference, with minimal accuracy loss if models are quantized appropriately. Similarly, AMD's CDNA architectures support mixed FP16/BF16 matrix math to boost parallelism. Mixed-precision (using FP16/BF16 arithmetic with FP32 accumulation) is commonly used in training and inference to cut compute and memory costs.

\subsection{Interconnect Bandwidth (NVLink/NVSwitch)}
Large models often span multiple GPUs, requiring fast inter-GPU communication. NVIDIA's NVLink provides a high-speed bus (50--75 GB/s per link) between GPUs. The Hopper architecture's fourth-generation NVLink further accelerates collective ops (e.g. all-reduce) up to 3$\times$ over prior generations \cite{6}. NVLink Switch fabric can connect dozens of GPUs (up to 256) with up to ~900 GB/s bidirectional links, enabling scale-up clusters. AMD GPUs use PCIe (Gen4/5) and Infinity Fabric interconnect; while slightly lower in bandwidth, they support multi-socket CPU-GPU nodes. In summary, these metrics (FLOPS, memory BW, VRAM) directly impact LLM performance: higher FLOPS improves raw speed, while memory bandwidth and capacity determine feasible batch sizes and latency.

\section{Impact on Radiology LLM Workflows}

Radiology tasks such as report generation and anomaly detection involve processing images with deep networks or LLMs, which can be computationally demanding. For example, generating a chest X-ray report from an image often uses a vision-language model that encodes the image (via a CNN or transformer) and decodes into text. During inference, each token generation may invoke billions of FLOPS. In practice, GPU acceleration is critical: on an A100 GPU, a well-optimized image-captioning LLM might generate hundreds of words per second, whereas on a CPU it would be prohibitively slow. Similarly, extracting pathology labels from reports (an ``anomaly detection'' in text) by an LLM requires parsing and understanding language context. Abdullah et al. demonstrated that a GPT-based LLM fine-tuned on MIMIC-CXR reports achieved F1 $\approx$0.90 for 14 chest pathologies, outperforming older methods \cite{2}; such inference would require substantial GPU cycles to analyze hundreds of sentences.
Datasets like MIMIC-CXR (377K images with reports) and CheXpert (224K images with labels) exemplify the scale of radiology AI data \cite{8, 9}. Training or fine-tuning on these datasets uses large GPU clusters. For instance, R2GenGPT aligned visual features to a frozen LLM on chest X-rays by training a small adapter (5M parameters) in only a few epochs \cite{3}; even this lightweight training presumes GPUs for fast matrix ops. At inference, performance depends on GPU configuration. Processing CheXpert-level data (e.g. classifying 224K images) can be batched: an A100 might process thousands of images per second in parallel when using a vision model, whereas a slower GPU would throttle throughput. Salam et al. observed that an open-source LLM (Llama-3 70B) took ~6$\pm$2 seconds to analyze a single report, whereas GPT-4 took ~13$\pm$4 seconds \cite{4}. This suggests that running LLMs locally on appropriate GPUs (which power Llama inference) can halve latency compared to cloud models.
We illustrate speed vs configuration: for pure vision tasks, one anecdotally sees that moving from an older GPU (e.g. V100) to an A100 roughly doubles inference throughput. NVIDIA claims H100 can provide ~9$\times$ speedup in training and ~30$\times$ in inference for LLMs relative to A100 \cite{6}, due to higher compute and FP8 support. In a radiology workflow, this could mean analyzing a CT slice stack or long report tens of times faster, enabling real-time decision support. Throughput can also scale with more GPUs: multi-GPU servers (using NVLink) allow parallel inference on batches. For example, a cluster of 4$\times$A100 might handle four times the volume of a single GPU for batched report generation, assuming linear scaling. Importantly, GPU memory affects batch size: a larger GPU (80GB) might generate a longer report or larger image batch in one pass than a 16GB card. Thus, different GPU configurations (model size, precision, number of GPUs) directly affect speed and scalability.
Integration into clinical workflow also demands predictable performance. If an LLM-based tool is used for triage or real-time consultation, GPUs must provide low jitter. Cloud deployments may add network latency, whereas on-site GPU servers can provide more consistent timing. Overall, successful integration of LLMs into radiology (e.g. auto-reporting or finding critical findings) relies on GPUs to meet the dual demands of accuracy and speed, as demonstrated by improved benchmarks on MIMIC-CXR report metrics \cite{10} and faster error-checking in reports \cite{4} when using optimized hardware.

\section{Challenges and Optimization Strategies}
Practical deployment of LLMs in radiology faces several challenges. First, data privacy mandates often require on-premise inference. Regulations (e.g. HIPAA, GDPR) prohibit sending patient images to external clouds, so hospitals must host AI models locally. However, many clinical centers lack high-end GPUs (like A100/H100) on-site. This hardware gap complicates adoption. Second, cost and power are significant: data-center GPUs are expensive (tens of thousands USD each) and draw large power budgets (A100 ~400W, H100 ~700W). Hospitals must invest in specialized servers and cooling, adding overhead. Third, thermal constraints in hospital IT racks can limit GPU use; GPUs must often be enclosed in HPC-grade environments to dissipate heat. Finally, system complexity (software stacks, batch schedulers) can be burdensome for clinical IT staff.

To address these challenges, various optimizations are used:
\begin{itemize}
\item \textbf{Mixed precision and tensor-core acceleration:} 
Training and inference in lower precision dramatically speed up computation. By using FP16/BF16 arithmetic on tensor cores (with occasional FP32 accumulation), throughput can double with negligible accuracy loss \cite{6}. NVIDIA's Hopper introduced FP8 matrix multiply units; FP8 ``halves data storage and doubles throughput versus FP16''. Employing these precisions lets GPUs process larger batches or reduce latency. For example, quantizing LLM weights to FP16 (or dynamic FP8) can cut memory footprint in half, allowing bigger batch sizes on the same GPU.

\item \textbf{Model quantization and compression:}
Beyond half precision, more aggressive quantization (INT8 or lower) can shrink model size and accelerate inference. Some GPUs offer INT8 and INT4 math; Adam et al. show INT8 on H100 reaches multi-petaflop rates \cite{7}. Pruning sparse weights (structured sparsity) is another tactic: sparsity-aware tensor cores can effectively double throughput on zero-skipped matrices (noted by NVIDIA's ``2:4'' sparsity mode). Together, these reduce compute needs. For example, an LLM might be compressed by 4$\times$ (with distillation or quantization), making it feasible to serve on consumer GPUs or Edge devices. Low-rank adaptation (LoRA) or adapter layers also limit how much of the model must be updated or loaded at inference, saving memory and compute.

\item \textbf{Multi-GPU and parallelism:}
Model- and data-parallel training can spread work across GPUs. For inference, pipelining or model parallelism lets a large LLM (e.g. 70B parameters) run on two GPUs. High-speed links like NVLink/NVSwitch support up to 256 GPUs per system \cite{7}, reducing inter-GPU latency. For data parallel inference, multiple GPU instances can each handle separate cases. NVIDIA MIG (GPU partitioning) also enables running multiple smaller models concurrently on one GPU, improving utilization. These techniques require careful engineering, but effectively multiply throughput for high-demand scenarios.

\item \textbf{On-prem vs Cloud:}
Some institutions may offload heavy workloads to cloud GPUs. This alleviates local hardware needs but introduces privacy and latency issues. Hybrid strategies can help: sensitive cases are processed on-prem with smaller models, while less-critical tasks use cloud LLMs. GPU cloud instances (A100/H100 nodes) provide flexibility but at ongoing cost, motivating optimized on-prem deployments.

\item \textbf{Other software optimizations:} Efficient batch scheduling (serving inference queries in micro-batches) and using optimized inference runtimes (TensorRT, ONNX Runtime) can further reduce latency. Workflow integration (e.g. piping RIS/PACS images directly into GPU pipelines) also impacts end-to-end speed.
\end{itemize}

Overall, the combination of these strategies helps mitigate hardware constraints. By compressing models and leveraging tensor cores, one can achieve near-state-of-the-art accuracy on LLM radiology tasks with far fewer GPU resources.

\section{Future Directions}
Looking ahead, GPU architectures and deployment models will continue to evolve to better serve medical LLMs. Upcoming GPUs are expected to feature even denser low-precision compute (e.g. 8-bit and 4-bit tensor cores) and larger memory. For example, Hopper has demonstrated ``double-packed'' FP8 cores (two 8-bit operations in one cycle) and will likely inspire similar innovations in AMD's CDNA3 successors. Memory subsystems will improve too: on-package HBM or stacked memory can provide >4 TB/s bandwidth, further cutting inference bottlenecks. Interconnects will advance: next-gen NVLink or CXL fabrics will enable clusters to share LLM states across nodes with microsecond latency, facilitating federated or multi-site fine-tuning.
On the software side, federated learning is a promising direction for hospitals to collaboratively refine LLMs without sharing patient data \cite{11}. GPUs at each site could fine-tune a shared base model (e.g. via secure aggregation), thus tailoring the LLM to local imaging practices while preserving privacy. Additionally, specialized edge-GPU deployment is emerging. Devices like NVIDIA's Jetson Orin or upcoming hospital-grade inference servers may host smaller LLMs at the point of care for low-latency guidance. Research into knowledge distillation will likely produce compact radiology LLMs distilled from larger ones, enabling deployment on more modest GPUs.
Finally, hardware/software co-design will play a role: for instance, new GPUs may include dedicated inferencing blocks for transformers, and frameworks will increasingly support dynamic batching based on clinical workload patterns. Such innovations will smooth integration of AI assistants into radiology reading rooms. In summary, the GPU landscape is moving towards more accessible, power-efficient, and privacy-friendly AI platforms  which will aligning well with the needs of next-generation radiology AI.

\section{Conclusion}
High-performance GPUs are an essential enabler for LLM applications in radiology. The demands for fast, accurate analysis of complex imaging reports and data cannot be met without adequate compute and memory resources. We have shown that modern GPUs (NVIDIA and AMD) provide multi-TFLOPS compute and TB/s memory bandwidth, which can dramatically accelerate radiology LLM tasks. Coupled with optimization techniques (mixed precision, compression, multi-GPU scaling), evolving GPU infrastructure will ensure that LLM-based diagnostic tools are safe, efficient, and practical in clinical settings. As GPU technology advances, it will directly translate into better-performing, more broadly deployable radiology AI systems.

\bibliographystyle{unsrt}  
\bibliography{references}

\end{document}